\documentclass[journal,10pt]{IEEEtran}
\ifCLASSOPTIONcompsoc
  \usepackage[nocompress]{cite}
\else
  \usepackage{cite}
\fi
\ifCLASSINFOpdf
\else
\fi
\usepackage{amssymb}
\usepackage[cmex10]{amsmath}
\usepackage{algorithmicx}

 \usepackage{dblfloatfix}
\usepackage{url}

\usepackage[normalem]{ulem}
\usepackage{gensymb}

\usepackage{amssymb}
\usepackage{booktabs}
\usepackage{url}
\usepackage[normalem]{ulem}
\usepackage{color}
\usepackage{adjustbox}
\usepackage{framed}
\newtheorem{theorem}{Theorem}
\newtheorem{corollary}{Corollary}
\newtheorem{lemma}{Lemma}
\usepackage{graphicx}
\usepackage{enumitem}
\usepackage{epstopdf}
\usepackage{booktabs}
\usepackage{multirow}
\usepackage{color}
\usepackage[numbers,square,comma]{natbib}
\usepackage{algorithm}
\usepackage{algpseudocode}

\newcommand{\T}{\tilde{T}}
\newcommand{\f}{\mathbf{f}}
\newcommand{\Tv}{\mathbf{T^*}}
\newcommand{\Pv}{\mathbf{P^*}}
\newcommand{\J}{\mathbf{J}}
\newcommand{\A}{\mathbf{A}}
\newcommand{\B}{\mathbf{B}}
\newcommand{\I}{\mathbf{I}}

\begin{document}

%
\title{Analysis and Control of Power-Temperature Dynamics in Heterogeneous Multiprocessors}
%
%
%
%

\author{Ganapati Bhat, Suat Gumussoy, Umit Y. Ogras

\thanks{
	 Ganapati Bhat and Umit Y. Ogras are with the School of Electrical, 
Computer and Energy Engineering, Arizona State University, Tempe, AZ. 
\protect\\
	E-mail: \{gmbhat, umit\}@asu.edu}
	\thanks{Suat Gumussoy is an IEEE Member.
	E-mail: suat@gumussoy.net} 
\thanks{ This article has been accepted for publication in IEEE Transactions on 
Control Systems Technology. DOI: 10.1109/TCST.2020.2974421}
\thanks{\copyright 2020 IEEE.  Personal use of this material is permitted.  
	Permission from IEEE must be obtained for all other uses, in any current or 
	future media, including reprinting/republishing this material for 
	advertising 
	or promotional purposes, creating new collective works, for resale or 
	redistribution to servers or lists, or reuse of any copyrighted component 
	of 
	this work in other works.}
}
\IEEEtitleabstractindextext{%
\begin{abstract}
Virtually all electronic systems try to optimize a fundamental trade-off 
between higher performance and lower power consumption. 
The latter becomes critical in mobile computing systems, such as smartphones, which rely on passive cooling.  
Otherwise, the heat concentrated in a small area drives both the junction and skin temperatures up. 
High junction temperatures degrade the reliability, while skin temperature 
deteriorates the user experience. 
Therefore, there is a strong need for a formal analysis of power consumption-temperature dynamics 
and predictive thermal management algorithms. 
This paper presents a theoretical power-temperature analysis of multiprocessor systems,
which are modeled as multi-input multi-output dynamic systems. 
We analyze the conditions under which the system converges to a stable steady-state temperature. 
Then, we use these models to design a control algorithm that manages the temperature 
of the system without affecting the performance of the application. 
Experiments on the Odroid-XU3 board show that the control algorithm is able to 
regulate 
the temperature with a minimal loss in performance when compared to the default thermal governors.


\end{abstract}

}

\maketitle

\IEEEdisplaynontitleabstractindextext

%
\IEEEpeerreviewmaketitle


%
%
%
\section{Introduction}
The performance and capabilities of state-of-the-art mobile processors are 
severely 
limited by heat dissipation and resulting chip 
temperature~\cite{huang2006hotspot,egilmez2015user}. 
Competitive performance and functionality are enabled by increasing the 
operating frequency and the number of processing 
cores~\cite{kumar2005heterogeneous}. These choices lead to higher device temperature, thus affecting both user experience and reliability~\cite{huang2006hotspot,egilmez2015user,note7recall}.
Therefore, maximum temperature constrains the power consumption,
which limits the maximum operating frequency and number of active 
cores~\cite{kumar2008system}. 
Commercial mobile platforms typically have hard-coded 
maximum safe temperature limits~\cite{pallipadi2006ondemand,pagani2017thermal}.
If the temperature exceeds this limit at runtime, the system throttles the operating frequencies or reduces the number of active cores
to decrease the temperature. 
In extreme cases, the system performs an emergency shutdown, which further 
cripples the performance and user experience~\cite{note7recall}. 

Power consumption and temperature form a well-known positive feedback 
system~\cite{vassighi2006thermal,liao2003coupled}. 
An increase in power consumption leads to an increase in temperature due to 
thermal resistance and capacitance networks~\cite{sharifi2013prometheus}. As a 
result of the increase in 
temperature, there is an exponential rise in the leakage 
power~\cite{roy2003leakage}.
If the system dynamics is stable, there exists a fixed-point temperature to 
which all the temperature trajectories converge within the region of convergence. That is, the temperature and power consumption rise until 
they eventually reach a fixed-point temperature.
In contrast, an unstable system leads to a thermal 
runaway~\cite{liao2003coupled,vassighi2006thermal,bhat2017power}.

The stability of the system and existence of a fixed point do not necessarily 
ensure a thermally safe 
operation~\cite{bhat2017power,liao2003coupled}. The 
stable fixed point may lie 
beyond the temperature at which the system starts to throttle the frequencies 
or it may be greater than the operating limit of the device.
Moreover, in a multiprocessor system, each core can converge to a different 
fixed-point temperature due to interactions between the core and 
workload running on each core~\cite{sahin2016qscale}.
For instance, if the graphics processing unit~(GPU) is not heavily utilized, it 
converges to a lower temperature than heavily utilized cores.
Therefore, the stable fixed point for each temperature hotspot has to be 
evaluated at runtime as the workload changes, so that necessary actions can 
be taken when violations are detected.

This paper presents a theoretical analysis of power-temperature 
dynamics in mobile multiprocessor systems. We start with an overview of power and thermal models 
used in the analysis.
Then, we present an efficient method to evaluate the stability of the system 
and compute 
its fixed points. This is followed by a detailed 
analysis of the region of convergence of the thermal dynamics in a mobile multiprocessor system. 
Finally, we present a control algorithm to illustrate how we can avoid thermal violations. 
The control algorithm uses the fixed-point 
predictions to determine if the power-temperature dynamics is within a safe 
region of operation. 
If any violations are detected, it reduces the frequency for the application that is causing 
thermal violation. The selective control ensures that other applications in 
the system are not penalized. 
We evaluate the control algorithm on the 
Odroid-XU3~\cite{ODROID_Platforms} platform. The platform includes the 
Exynos-5422 big.Little 
System-on-a-Chip~(SoC). We use this platform for our experimental evaluations 
since it is used on smartphones like Samsung Galaxy S5 and is representative of 
the processors used in mobile systems. Our experiments on the
Odroid-XU3 platform show that the proposed algorithm is 
able to isolate the power-hungry application, while maintaining the performance 
requirements of active applications. 
In summary, the major contributions of this paper are:

\begin{itemize}
	\item An efficient solution to find the region of convergence of 
	the power-temperature dynamics in a multiprocessor system,
	\item An optimization technique that speeds up the computation of 
	the fixed 	point by up to 1.8$\times$ when compared to a baseline approach used 
	in~\cite{bhat2017power},
	\item A low-overhead control algorithm that uses fixed-point 
	analysis to prevent thermal violations and lower the power consumption. 
	Implementation on a commercial 
	device shows that the algorithm is effective in preventing thermal 
	violations while also lowering the power consumption of the device.
\end{itemize}

\noindent\textbf{Application domains of proposed analysis:}
The primary application area of the proposed analysis is low-power handheld and wearable devices where the skin temperature of the device is critical for user comfort. This is especially important in wearable devices where the device is in constant contact with the user's body.  We can use the proposed analysis to predict the steady-state temperature of the device when a user is running an application over long periods of time (such as playing a game). Then, we can use the predictions and the time to reach the steady state to determine if the current trajectory leads to high skin-temperature and user discomfort. We also note that the proposed analysis and the control algorithm are not directly applicable to high-performance server units where the power consumption may change in the order of milliseconds.

\vspace{1mm}
The rest of the paper is organized as follows. We review the related work in 
Section~\ref{sec:related_work}. Section~\ref{sec:overview} presents an overview 
of the 
temperature and power models used in the fixed-point analysis.
Section~\ref{sec:fixed_point_prediction} first presents an overview of the 
fixed-point prediction for a single-input-single-output~(SISO) model. This is 
followed 
by our analysis of the general multiple-input-multiple-output~(MIMO) model. The 
lightweight control algorithm using fixed-point predictions is presented in 
Section~\ref{sec:control_algorithm}. Finally, we present the experimental 
evaluations in Section~\ref{sec:experiments} and the conclusions in 
Section~\ref{sec:conclusion}.

\section{Related Work} \label{sec:related_work}
Higher power densities in modern multiprocessors has led to increasing thermal 
stress on these devices. Higher temperature has an adverse impact on 
reliability~\cite{vassighi2006thermal,li2004efficient}.
Therefore, a significant amount of research has focused on accurate, yet 
complex, offline models and lightweight runtime models. 
Design-time studies focus on analyzing the full-chip thermal 
behavior under different application workloads such that thermal policies for 
the chip can 
be determined~\cite{huang2006hotspot,zhan2005high}.
Since they simulate traces from common workloads to model the 
static and transient thermal behavior, they are not amenable to runtime analysis.
In contrast, runtime approaches develop 
computationally efficient thermal and power models, which can be used for 
runtime analysis of the 
system~\cite{beneventi2014effective,bhat2017power}. 
For instance, a 
grey-box system identification is used in~\cite{beneventi2014effective} to 
develop the thermal model of a multicore system. A recent approach proposed 
in~\cite{bhat2017power} performs an analysis of the stability of 
power-temperature dynamics in multiprocessor systems. However, it uses a 
simplified SISO model for evaluating the region of convergence of the 
power-temperature dynamics. In contrast, this paper presents an efficient 
solution to find the region of convergence of the power-temperature dynamics 
using a MIMO model.

MIMO system identification techniques have also been used in dynamic power management of multiprocessor systems~\cite{ogras2009design,bogdan2015energy}. For instance, the work in~\cite{ogras2009design} uses a MIMO formulation to model the queue utilizations in a multiprocessor system with multiple voltage-frequency islands. Then, it uses a feedback control algorithm to determine the operating frequency and voltage of each island while maintaining a reference utilization for each queue. Similarly, the approach in~\cite{bogdan2015energy} first models the workload characteristics in a multiprocessor system using a multi-fractal master equation. Then, it uses the equation in a model predictive control algorithm to perform dynamic voltage and frequency scaling. Complementary to these approaches, the MIMO system identification used in this paper models the temperature of the device as a function of the power consumption of the different components in the system while not considering the specific workload characteristics. Therefore, we can potentially combine our models with the MIMO approaches of prior work~\cite{ogras2009design,bogdan2015energy} to enable workload-aware thermal and power management.

Commercial products come with a mechanism to control the 
temperature of the system. It is achieved either by throttling the performance 
or shutting down the system. Reactive approaches continuously monitor the 
temperature and respond to avoid thermal violations, resulting in a larger 
penalty on performance~\cite{sahin2016qscale,bhat2019power}.
To mitigate the performance penalty, recent research has focused on developing 
predictive approaches for dynamic thermal and power management~(DTPM).
Predictive approaches develop computationally efficient thermal and power 
models~\cite{beneventi2014effective,bhat2017algorithmic}.
These models are used at runtime to estimate the short-term behavior of the 
power-temperature dynamics and take actions if violations are 
predicted~\cite{cochran2013thermal,	singla2015predictive, bhat2017algorithmic}.
These approaches are effective in predicting the short-term behavior of the 
power-temperature dynamics; however, the prediction error increases 
when long-term predictions are made~\cite{singla2015predictive}. 
A number of studies have also used control-theoretic approaches to regulate the temperature of the system~\cite{rao2015temperature,zanini2009control,leva2018event,mutapcic2009processor}. 
For instance, the work in~\cite{rao2015temperature} uses an integral controller to regulate the temperature of the system. 
The authors consider a nonlinear thermal model that captures the exponential dependence on leakage power. 
Similarly, the authors in~\cite{leva2018event} use an event-based 
proportional integral controller to maintain the temperature of the system below a desired set point. 
These approaches implement control algorithms around a desired set point. If the temperature reaches this set point, they throttle the whole system to decrease the power consumption and regulate the temperature. This, however, can cripple the whole system performance and penalize all active applications. In contrast, our stability analysis technique predicts the steady-state temperature before a thermal violation occurs. Hence, it can assist thermal management by predicting the steady-state temperature and enabling proactive decisions.


Positive feedback between the power consumption and temperature has been analyzed by a number of 
studies~\cite{liao2003coupled,heo2003reducing,bhat2017power}.
For instance, Liao et al.~\cite{liao2003coupled} show that a positive 
second-order derivative of the temperature leads to a thermal runaway. This criterion can be 
easily used for design-time analysis of the chip. However, it is not practical 
to use it at runtime, since the condition is true only after a thermal runaway has started.
Temperature dependence of leakage current is 
used in~\cite{heo2003reducing} to increase the thermally sustainable power of the chip.
The work in~\cite{bhat2017power} proposes an approach to analyze the existence 
of fixed points by first reducing the system to a SISO model.
However, none of these techniques can analyze the existence and stability of fixed points 
at runtime while considering the coupling between different cores. 
This paper presents a theoretical methodology to overcome this limitation and provides an efficient 
algorithm to calculate the fixed points and their stability. The fixed points 
are then used in a simple control algorithm that can avoid thermal violations 
without excessive performance penalties. 
The fixed-point analysis can also be used as an effective guard against power attacks, 
which cause damage by increasing the temperature of the device~\cite{dadvar2005potential,hasan2005heat,kong2010thermal}.

\section{Background and Overview} \label{sec:overview}
This section presents the power consumption and temperature models
required for the proposed thermal fixed-point analysis. We provide an overview 
here for completeness, while additional details are presented 
in~\cite{bhat2017algorithmic,bhat2017power}.

\subsection{Power and Temperature Models} \label{sec:models}
Let us assume that there are $M$ processors in the system, as summarized in 
Table~\ref{tab:summary_symbols}.
The power consumption of processors in the target 
system is given by the $M \times 1$ vector $\mathbf{P} = [P_1, P_2, \ldots, 
P_M]^\text{T}$,
where $P_i$ is given by:
%
\begin{equation} \label{eqn:total_power}
P_i = C_{sw,i} V_{i}^2 f_i + V_{i}I_{g,i} +  V_{i}\kappa_{1,i} T_i^2 
e^\frac{\kappa_{2,i}}{T_i},~~1 \leq i \leq M.
\end{equation}
where $C_{sw,i}$ is the switching capacitance, $V_i$ is the supply voltage and 
$f_i$ is the operating frequency, $I_{g,i}$ is the gate leakage, $\kappa_{1,i} 
>0$ and $\kappa_{2,i}< 0$ are technology-dependent parameters, and $T_i$ is the 
temperature. Since our focus is to analyze the dependence of power on the 
temperature, we can isolate the impact of the temperature as
\begin{equation}\label{eqn:power_simplifed}
P_i = P_{C,i} + V_i \kappa_{1,i}T_i^2e^{\frac{\kappa_{2,i}}{T_i}}, ~~1 \leq i 
\leq M,
\end{equation}
where $P_{C,i} = C_{sw,i} V_i^2 f_i + V_i I_{g,i}$ represents the 
\emph{temperature-independent} component.

Assume that are $N$ thermal hotspots in the system.
The temperature of each thermal hotspot can be written as a function of the 
power 
consumption vector $\mathbf{P}$ and
thermal capacitance and conductance matrices~\cite{huang2006hotspot,sharifi2013prometheus}.
Therefore, the temperature dynamics is defined by the state-space system
\begin{equation}\label{eqn:temp_model}
\mathbf{T[k+1]} = \A \mathbf{T[k]}+ \B \mathbf{P[k]}
\end{equation}
where $\A$ matrix is symmetric due to symmetricity of the heat transfer.
A discrete-time system is used because temperature measurements and control 
decisions are sampled at equal discrete time intervals. 
Equation~\eqref{eqn:temp_model} expresses the temperature of each hotspot in 
the 
next time step as a function of the current temperature and the power 
consumption of $M$ sources. 
The $\A$ matrix describes the effect of temperature in time step $k$ 
on the temperature in the next time step. That is, it describes the coupling 
between the temperature hotspots.
Similarly, the ${N \times M}$ matrix $\B$ describes the 
relation between the temperature in the next time step and each of the power 
sources.
Substituting~\eqref{eqn:total_power} in~\eqref{eqn:temp_model} leads to a 
system of nonlinear equations,
\begin{align} \label{eqn:nonlinear_model}
\mathbf{T[k+1]} =\A\mathbf{T[k]} +
\B \big[P_1[k], P_2[k], \ldots, P_M[k] \hspace{0mm}
	\big]^\text{T},~\mathrm{where} \\ \vspace{-2mm}
P_i[k] =C_{sw,i} V_{i}^2 f_i + I_{g,i} V_{i} +  V_i\kappa_{1,i} T_i[k]^2
	e^\frac{\kappa_{2,i}}{T_i[k]},
	\hspace{-0.25mm}~1 \hspace{-0.25mm} \leq \hspace{-0.25mm}i
	\hspace{-0.25mm}\leq \hspace{-0.25mm} M \nonumber.
\end{align}
%
\begin{table}[t]
	\centering
	\small
	\caption{Summary of major parameters}
	\label{tab:summary_symbols}
	\begin{tabular}{@{}ll@{}}
		\toprule
		Symbol                      & 
		Description                                                             
		
		\\
		\midrule
		$N,M$                         & \begin{tabular}[c]{@{}l@{}} The number 
			of 
			thermal hotspots and processing \\
			elements (resources) in the SoC, respectively. 
		\end{tabular}                                                       
		\\
		\midrule
		$\mathbf{T[k]}$                      & \begin{tabular}[c]{@{}l@{}}$N 
			\times 1$ 
			vector where $T_i[k],~1 \leq j \leq N$\\ denotes the temperature of 
			the 
			$i^{th}$ hotspot \\ at time instant k.\end{tabular}        \\ \midrule
		$T$                         & \begin{tabular}[c]{@{}l@{}} 
			The maximum (scalar) steady-state temperature \\
			over all thermal hotspots.\end{tabular}                       \\ 
		\midrule
		$\mathbf{P[k]}$                      & \begin{tabular}[c]{@{}l@{}}$M 
		\times 1$ 
			vector where $P_i[k],~1 \leq i \leq M$ denotes\\ the power 
			consumption 
			of the $i^{th}$ resource \\ at time instant k.\end{tabular} \\ \midrule
		$\kappa_{1,i}, \kappa_{2,i}$ & 
		\begin{tabular}[c]{@{}l@{}}Technology-dependent parameters 
		of the\\ leakage power for the $i^{th}$ 
			resource \\ where $\kappa_{1,i} 
>0$ and $\kappa_{2,i}< 0$.\end{tabular}                            \\ \midrule
		$a, b$                        & \begin{tabular}[c]{@{}l@{}}Parameters 
		of 
			the single input single output\\ model which describes the thermal 
			dynamics.\end{tabular}  \\ \midrule
		$\A, \B$                        & \begin{tabular}[c]{@{}l@{}}Parameters 
			of 
			the multiple input multiple output\\ model which describes the 
			thermal 
			dynamics.\end{tabular}  \\ \midrule
		$\T, \alpha, \beta$   & Auxiliary parameters introduced in 
		Equation~\ref{change_of_parameters}. 
		\\ 	\bottomrule
	\end{tabular}
\end{table}
The nonlinear system of equations in~\eqref{eqn:nonlinear_model} captures the 
positive feedback between power consumption and temperature. 
The steady-state solution of this system gives the fixed points of the thermal hotspots in the system. The steady-state temperature and power consumption are functions of dynamic changes in the power consumption, operating frequency of the cores, and the ambient temperature. Therefore, the steady-state solution of \eqref{eqn:nonlinear_model} has to be evaluated at runtime.
Evaluating the steady-state solution of \eqref{eqn:nonlinear_model} at runtime is difficult due to the following reasons:
\begin{enumerate}
	\item The system may not be stable due to the positive feedback between 
	power and temperature,
	\item The power consumption and temperature at any time instance depend on 
	their previous values. This necessitates an iterative approach, which is 
	not suitable for runtime systems~\cite{bhat2017power},
	\item The region of convergence of the system in terms of temperature and 
	power consumption of each resource has to be found to ensure thermally safe 
	operation.
\end{enumerate}

In order to address these challenges, we first reduce the MIMO system in \eqref{eqn:nonlinear_model} to a SISO system and then find the steady-state solution for the SISO system. Using the SISO solution of each temperature hotspot we obtain the steady-state solution of \eqref{eqn:nonlinear_model} using Newton's method, as described in the rest of this section.
\subsection{Background on Fixed-Point Analysis of the SISO System}

In this section, we summarize our fixed-point analysis for the reduced-order SISO system 
since it provides the necessary background for the MIMO analysis. 
The details of the reduced order SISO analysis can be found in~\cite{bhat2017power}. 
The MIMO power-temperature dynamics can be reduced to a SISO model for each of the temperature hotspots.
We start with a reduced-order SISO system because it allows us 
to do an in-depth theoretical analysis of the system. 
To this end, we identify the parameters of the reduced-order system 
through system identification. 
Since the thermal safety is determined by the maximum 
temperature, the following analysis is performed for the maximum temperature.

The maximum steady-state temperature across all the hotspots can be denoted by a scalar
$T$ as:
\begin{equation*}
T = \max_{1 \leq i \leq N} \lim_{k \rightarrow \infty} T_i[k]
\end{equation*}
At steady state ($k\rightarrow \infty$), the temperature of each hotspot can be 
modeled as, 
\begin{equation}\label{eqn:temp_ss}
T = aT + bP,
\end{equation}
\noindent where $T$ is the steady-state temperature, $P$ is the steady-state power consumption and $0<a<1$, $b>0$ are the parameters of the reduced-order SISO 
system.
Using~\eqref{eqn:power_simplifed} in~\eqref{eqn:temp_ss} gives
\begin{equation} \label{fixed_point_ori}
(1-a)T - bP_C = bV \kappa_{1}T^2e^{\frac{\kappa_{2}}{T}}.
\end{equation}
The power-temperature dynamics has fixed points if this equation has solutions.
This means that if the dynamic power consumption of the device is maintained at $P_C$, the device will attain a steady state that is given by the solution of~\eqref{fixed_point_ori}. Conversely, the dynamics does not have fixed points either when \eqref{fixed_point_ori} does not have a solution or continuous variations in the power consumption prevent the system from reaching a steady state.
Since we use \eqref{fixed_point_ori} to perform the stability analysis, the following 
change of variables is introduced to simplify the notation.
\begin{equation}\label{t_tilde}
\T = -\frac{\kappa_2}{T}, \hspace{3mm} 
\end{equation}
\begin{equation} \label{change_of_parameters}
\alpha = \frac{b}{a-1}\frac{P_C}{\kappa_2} > 0, \hspace{3mm}
\beta =  \frac{a-1}{b} \frac{1}{V \kappa_{1} \kappa_{2}} > 0 \hspace{1mm}.
\end{equation}
With this change of variables,~\eqref{fixed_point_ori} becomes
\begin{equation} \label{eqn:fixed_point}
\beta \T(1-\alpha\T)=e^{-\T},
\end{equation}
where $\alpha >0, \beta >0$. Next, we present the conditions on the auxiliary 
temperature $\T$, $\alpha$, and $\beta$ to ensure that \eqref{eqn:fixed_point} 
has a solution.


\subsubsection{Necessary and Sufficient Conditions for the Existence of Fixed 
	Point(s)} \label{sec:fixed_point_condition}
\begin{figure}[t]
	\centering
	\includegraphics[width=1\linewidth]{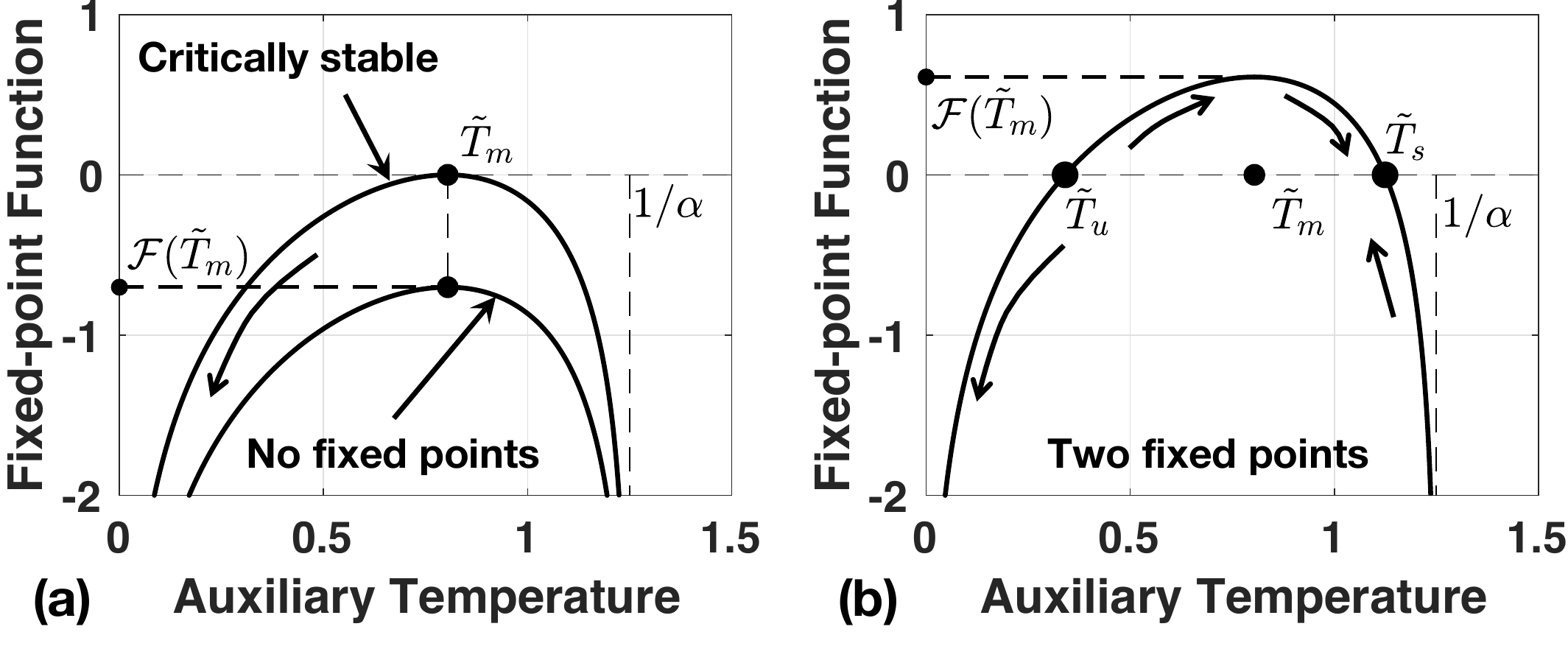}
	\caption{Illustration of fixed point function when there is (a) no fixed point and 
		(b) when there are two fixed points, respectively. The figure is adapted from~\cite{bhat2017power}.}
	\label{fixed_point_illustration}
	\vspace{-3mm}
\end{figure}

The domain of the auxiliary temperature in~\eqref{t_tilde} is given by $\T \in 
(0,\infty)$, 
since $\T = -\kappa_2 /T$ where $\kappa_2 < 0$. 
Hence, the right-hand side of~\eqref{eqn:fixed_point} 
lies in the interval $(0,1)$. 
That is, $0 < \beta \T(1-\alpha\T) = e^{-\T} < 1$ and $\T<1/\alpha$. 
Therefore, the logarithm of both sides can be taken while maintaining the 
equality, that is,
\[
\ln \beta + \ln \T + \ln(1-\alpha \T) = -\T.
\]
Therefore,~\eqref{eqn:fixed_point} has the same fixed points as,
\begin{equation} \label{eqn:fixed_point_ln}
{F}(\T) = \ln \beta + \ln \T + \ln(1-\alpha \T) + \T = 0.
\end{equation}
%
The following lemma 
from \cite{bhat2017power}
summarizes the properties of $\mathcal{F}(\T)$.
\begin{lemma}  \label{lemma_F_T}
	$\mathcal{F}(\T)$ given in Equation~\ref{eqn:fixed_point_ln} satisfies the 
	following properties:
	\begin{enumerate}
		\item $\mathcal{F}(\T)$ is a concave function in the interval $\T \in 
		(0,1/\alpha)$.
		\item $\mathcal{F}(\T)$ has a unique maxima at $\T_m$, which is given 
		by:
		\begin{equation} \label{eqn:inflection_point}
		\T_m = \frac{1}{2\alpha}-1+\sqrt{\frac{1}{4\alpha^2}+1}
		\end{equation}
		\item $\mathcal{F}(\T)$ is an increasing function in the interval 
		$(0,\T_m)$ and 
		a decreasing function in $(\T_m,1/\alpha)$.
	\end{enumerate}
\end{lemma}
An illustration of $\mathcal{F}(\T)$ is plotted in 
Figure~\ref{fixed_point_illustration}. It can be seen that the function is 
concave in $(0,1/\alpha)$. The concavity also leads to $\mathcal{F}(\T)$ 
increasing in $(0,\T_m)$ and decreasing in $(\T_m, 1/\alpha)$.
The maxima of  $\mathcal{F}(\T)$ can be evaluated using the maxima $\T_m$ 
defined in Lemma~\ref{lemma_F_T}. This is summarized in Corollary~\ref{thm_fixed_point} below.
%
%
\begin{corollary} \label{thm_fixed_point} 
	The maxima of $\mathcal{F}(\T)$ is given by:
	\vspace{-1mm}
	\begin{equation*} 
	\mathcal{F}(\T_m) = \ln \beta - \ln\left(\frac{2}{\T_m}+1 \right)e^{-\T_m}
	\end{equation*}
	Moreover, equation~\eqref{eqn:fixed_point} has two fixed points if the 
	maxima 
	of $\mathcal{F}(\T)$ is greater than zero. 
	Consequently,~\eqref{eqn:fixed_point} 
	has two fixed points if and only if
	\begin{equation} \label{eqn:F_Tm}
	\beta\geq\left(\frac{2}{\T_m}+1\right)e^{-\T_m}
	\end{equation}
	where $\T_m$
	depends only on the parameter $\alpha$ and it is defined 
	in~\eqref{eqn:inflection_point}.
	Otherwise, it has no solution.
\end{corollary}
%
At runtime, the value of $\T_m$ is computed using~\eqref{eqn:inflection_point}. 
Then, the condition for existence of fixed points given in Corollary~\ref{thm_fixed_point} (Eq. \ref{eqn:F_Tm}) is evaluated. 
If it is not satisfied, there will be a thermal runaway as presented in Theorem~\ref{thm_stability} below. 
In practice, this means that the system will either throttle cores aggressively 
or perform 
an emergency shutdown. 
\subsubsection{Stability of the Fixed Points}  \label{sec:stability}
The behavior of $\T$ as a function of $\mathcal{F}(\T)$ determines the stability 
of the fixed points. 
This behavior is summarized using the following lemma in \cite{bhat2017power}.
\begin{lemma} \label{lem_signFT}
	The value of $\T$ in the temperature iteration increases when 
	$\mathcal{F}(\T)<0$, 
	and decreases when $\mathcal{F}(\T)>0$.
\end{lemma}
This lemma can be used to determine the stability using the sign of 
$\mathcal{F}(\T)$. The following theorem in \cite{bhat2017power} gives the stability of the 
fixed points using this lemma.

\begin{theorem} \label{thm_stability}
	The stability of the fixed points is as follows:
	\begin{enumerate}
		\item When~\eqref{eqn:fixed_point_ln} has no solution,
		the temperature iteration diverges, i.e., $\T \rightarrow 0$ $(T 
		\rightarrow 
		\infty)$, 
		as illustrated in Figure~\ref{fixed_point_illustration}.
		Hence, there is a thermal runaway.
		\item When there are two fixed points, $\T_u\in(0,\T_m)$ is unstable 
		and $\T_s\in(\T_m,\frac{1}{\alpha})$ is stable.
		\item In the latter case, any temperature iteration starting 
		$(0,\T_u)$  
		diverges, i.e., $\T \rightarrow 0$ and $T \rightarrow \infty$ leading 
		to a 
		thermal runaway.
		However, any temperature iteration starting in 
		$(\T_u,\frac{1}{\alpha})$ 
		converges to the stable fixed point $\T_s$.
	\end{enumerate}
\end{theorem}
Theorem~\ref{thm_stability} states that the temperature 
proceeds along the arrows shown in Figure~\ref{fixed_point_illustration}.
When $\mathcal{F}(\T_m) < 0$, i.e., no fixed point exists, 
$\T$ decreases at each temperature iteration no matter where it starts.
This means that $T$ increases in each temperature iteration due 
to~\eqref{t_tilde}.
Therefore, there is a thermal runaway as illustrated in 
Figure~\ref{fixed_point_illustration}(a). 
When $\mathcal{F}(\T_m) > 0$, there are two fixed points denoted by $\T_u$ and 
$\T_s$. 
Iterations that start in the interval $(0,\T_u)$ diverge to $\T \rightarrow 
0$, 
since $\mathcal{F}(\T) < 0$ in $(0,\T_u)$.
Conversely, temperature iterations with initial points in the interval $(\T_u, 
\T_s)$ will converge to $\T_s$, since $\mathcal{F}(\T) > 0$. 
That is, temperature iterations starting in the interval 
$(\T_u,\frac{1}{\alpha})$ 
will also converge to $\T_s$, as illustrated in 
Figure~\ref{fixed_point_illustration}(b). 
Therefore, iterations starting in $\T_u\in(0,\T_m)$ are divergent,  
while iterations starting in $\T_s\in(\T_m,\frac{1}{\alpha})$ are convergent.

%
\section{Fixed-Point Analysis for MIMO System} 
\label{sec:fixed_point_prediction}
The SISO model in the previous sections allows us to analyze the behavior of 
thermal hotspots individually. However, in a multiprocessor system there is a 
close interaction of different resources in the system. As a result of this, 
activity in one part of the system affects the fixed-point temperatures in 
other parts of the system. Therefore, it is important to solve the MIMO system 
in~\eqref{eqn:nonlinear_model} to 
calculate the fixed-point temperatures. 
The steady-state equation of the temperature dynamics can be modeled as
\begin{equation}\label{eqn:mimo_steady_state}
\Tv = \A\Tv + \B\Pv
\end{equation} where $\Tv$ and $\Pv$ are the steady-state temperatures and power consumptions, respectively.
Define the following function,
\begin{equation}\label{eqn:temp_miso}
\mathbf{f(\Tv)}= (\A - \I)\Tv + \B\Pv. 
\end{equation}
where $\mathbf{f(\Tv)}$ is obtained by grouping the temperature terms in 
\eqref{eqn:mimo_steady_state}.
$\mathbf{f(\Tv)}$ is a vector-valued function that depends on $\Tv=[T_1, \ldots, 
T_N]^T$ and $\Pv=[P_1, \ldots, P_M]^T$ satisfying a set of nonlinear equations in terms of $\Tv$.
This is a nonlinear system of equations due to the exponential temperature 
terms in $P_i$. 
The standard approach to solving this problem is to find a good initial point and use a root-finding algorithm to 
solve the nonlinear equations. 
In this work, we use the Newton's method~\cite{atkinson2008introduction} to solve the system of 
equations in~\eqref{eqn:temp_miso}.
The number of iterations required for Newton's method depends on the 
initial points. An effective method to find good initial points is to use a 
reduced-order SISO system as described in~\cite{bhat2017power}.
The SISO solution provides fixed points with high accuracy and very low 
computational cost. Therefore, the SISO model of each hotspot can be used as 
the initial point in the Newton iterations. The convergence of Newton 
iterations, i.e., the existence of fixed points, will 
depend on the system parameters and the power consumption of each of the 
components. Next, we analyze the region of convergence of the power-temperature dynamics on the Odroid-XU3 board.

\subsection{Region of Convergence for Power-Temperature Dynamics}
\noindent The fixed-point function is given in~\eqref{eqn:temp_miso} as
\begin{equation}\label{eq:mimo_eval}
\f(\Tv) = (\A - \I)\Tv + \B\Pv.
\end{equation}
We can write the Jacobian matrix of $\f(\Tv)$ as
\begin{equation}
\J_\f(\Tv) = (\A-\I) + \B\Pv',
\end{equation}
where $\Pv'$ is the derivative of the power consumption values with respect to 
the temperatures. In each iteration of Newton's method, the temperature vector step is given by,
\begin{equation}\label{eqn:dt}
\Delta\Tv_{k} = -\J^{-1}_\f(\Tv_{k})\f(\Tv_{k})
\end{equation}
where $k$ denotes the current iteration of Newton's method. Using 
$\Delta\Tv_{k}$, the temperature vector is 
updated as,
\begin{equation}\label{eqn:temp_update}
\Tv_{k+1} = \Tv_{k} -\J^{-1}_\f(\Tv_{k})\f(\Tv_{k}),
\end{equation}
where subscripts $k$ and $k+1$ are used to denote temperature in the current 
iteration and the next iteration, respectively. Using~\eqref{eqn:temp_update}, 
we can define the Newton function $\mathbf{g}(\Tv_{k})$ as
\begin{equation}\label{eq:newton_function}
\mathbf{g}(\Tv_{k}) = \Tv_k + \Delta\Tv_k.
\end{equation}
The Newton function gives all the temperatures that will be encountered during 
the evaluations of Newton's method. We use this definition of the Newton function to analyze the stability and region of convergence of the power-temperature dynamics of the system.
Specifically, the system has a 
guaranteed convergence to the unique fixed point by Contraction Mapping  
Theorem (see p.220 in~\cite{rudin1976principles}) if the following conditions are satisfied: 

\begin{itemize}
	\item Range of the Newton function is 
	contained in the domain of the function. If the domain of the function is 
	defined as the safe operating limits of the device, this condition ensures 
	that all the temperature vectors given by Newton's method lie within 
	safe operating limits of the device.
	\item Norm of the Jacobian of the Newton function $\mathbf{g}(\Tv_{k})$ is 
	less than 1.
\end{itemize}

\noindent These conditions will be satisfied for a range of power consumption 
and 
system parameters. These bounds can be found by analyzing the properties of 
$\J^{-1}_\f(\Tv)$ and $\f(\Tv)$, which is challenging due to the non-linearity.
Therefore, in order to evaluate the region of convergence of the system, we analyze whether the 
previous conditions are satisfied for a range of power consumption values.

\begin{figure}[b]
	\centering
	\includegraphics[width=0.99\linewidth]{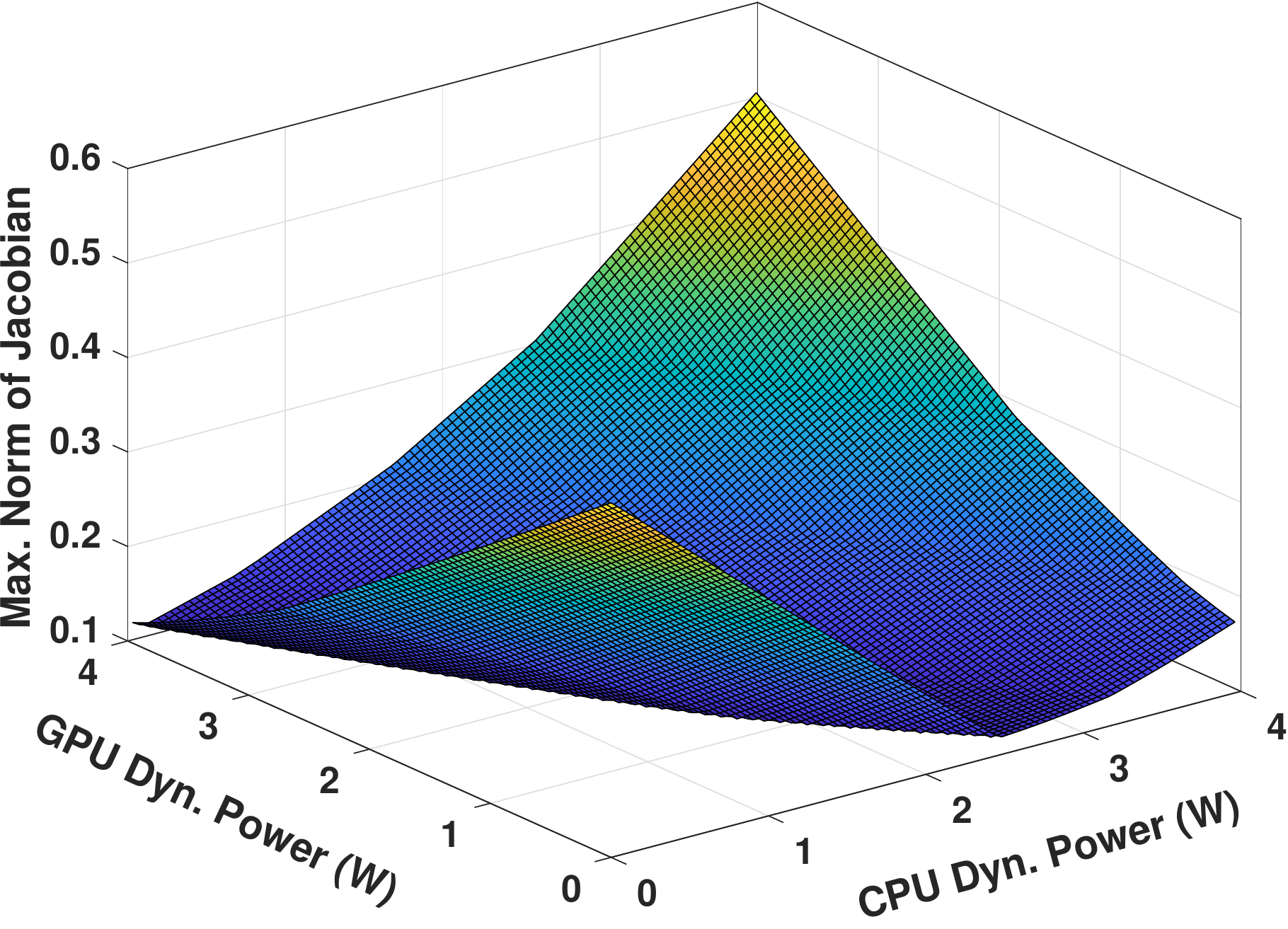}
	\caption{Maximum norm of the derivative of the Newton function when varying 
		the CPU and GPU power consumption.}
	\label{fig:norm_derivative}
	\vspace{-1mm}
\end{figure}

In order to evaluate the region of convergence of the power-temperature dynamics, 
we recursively evaluate the MIMO model equations~\eqref{eq:mimo_eval} 
to~\eqref{eq:newton_function}.
The system parameters that affect the region of convergence include $\A$, $\B$ matrices and the leakage power 
parameter of each resource. For a given set of system parameters, the 
power-temperature dynamics has a guaranteed convergence to a stable and safe fixed point for a range of power 
consumption and temperature values. The range of temperature for which the 
processor can operate safely is typically determined by the manufacturer. 
Therefore, the analysis performed in this section focuses on the range of 
power consumption for which convergence is guaranteed. As a first step, the 
nominal values of the system parameters determined from system 
identification are used to perform the simulations. Then, power consumption 
of the big cluster and GPU, which are the resources with temperature hotspots 
in the system, are swept from 0.0024 W to 4 W. We chose this range of power 
consumption values, since it is the 
typical power consumption range observed in a wide range 
of benchmarks. For each power consumption pair, the range of the Newton 
function $\mathbf{g(\Tv)}$ and derivative of the Newton function are evaluated 
over the allowable temperature range of 37$^\circ$C to 120$^\circ$C. 
Using these results, the range of power consumption values for which the power-temperature dynamics in the system converges to a safe temperature is determined.

\begin{figure}[t]
	\centering
	\includegraphics[width=1\linewidth]{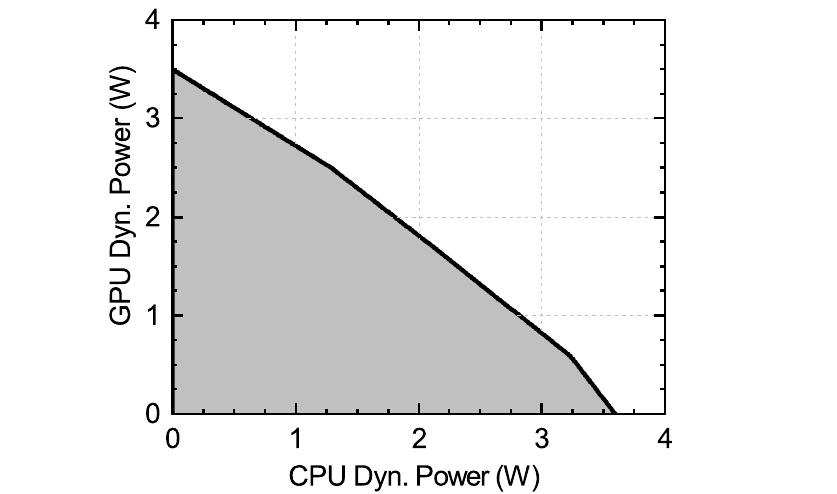}
	\vspace{-3mm}
	\caption{The region of power consumption values for which the range 
			of the Newton function is contained in the domain. 
			The proposed technique guarantees convergence when the total dynamic power consumption of 
			the CPU and GPU is less than about 3.5 W. 
			This result is aligned with our experiments on the Odroid-XU3 board where sustained operation at power consumption 
			higher than 3.5~W lead to thermal throttling of the system.}
	\label{fig:convergence}
	\vspace{-0.07in}
\end{figure}

In order to have guaranteed convergence, the norm of the derivative must be 
less than 1 and range of the Newton function must be contained in the domain. 
As we can see in Figure~\ref{fig:norm_derivative}, the maximum norm of the 
Jacobian 
is less than 1 for the entire range of power consumption values. However, 
when 
the CPU power is greater 3.5 W, the range of the Newton function 
$\mathbf{g}(\Tv_{k})$ is 
no longer contained in the domain. This means that there exists a temperature trajectory exceeding the maximum safe temperature of the board.
Similarly, if the GPU power consumption is greater than 3.5 W, the range of the 
Newton function is not contained in the domain, as shown in 
Figure~\ref{fig:convergence}.
Therefore, we can conclude that the system has guaranteed convergence when 
the power consumption of the CPU and GPU are in the shaded region shown in 
Figure~\ref{fig:convergence}. We see that the system has guaranteed convergence 
when the total dynamic power consumption of the CPU and GPU is less than about 
3.5 W. This agrees with our experiments on the Odroid-XU3 board where sustained 
operation at power consumption higher than 3.5~W lead to thermal throttling of the system.

\subsection{Accelerating Newton Iterations}
Newton's method involves the calculation of $\J^{-1}_\f(\Tv_{k})$ in each iteration. 
This can be computationally expensive since we have to perform the inversion for multiple Newton iterations every time the fixed points are evaluated. 
Therefore, we exploit the structure of the temperature dynamics of the experimental platform 
to speed up the Newton's method. 
The Odroid XU3 board consists of a little CPU cluster, a big CPU 
cluster, main memory and a GPU. Each CPU cluster contains four physical cores. 
The temperature hotspots in this system include the big cores and the GPU. 
Using this structure, the fixed-point function can be simplified by 
separating the temperature-dependent and temperature-independent terms as, 
\begin{equation}\label{eq:f_simplified}
\f(\Tv) = (\A-\I)\Tv+\B \begin{bmatrix}
p_{little} \\
p_{big,1} + p_{big,2} s_{big} \\
p_{mem} \\
p_{gpu,1} + p_{gpu,2} s_{gpu}
\end{bmatrix},
\end{equation}
where $p_{big,1}$ and $p_{gpu,1}$ are temperature-independent components of the 
big cluster and GPU, respectively. Temperature-dependent components of only the 
big cluster and GPU are considered as they have the largest effect on the 
temperature hotspots. The temperature-dependent components of 
power are defined as,
\begin{align} \nonumber
p_{big,2}& =V_{big}\kappa_{1,big},& s_{big}& 
=T_{big}^2\exp\Big(\frac{-\kappa_{2,big}}{T_{big}}\Big), \\ \nonumber
p_{gpu,2}& =V_{gpu}\kappa_{1,gpu},& s_{gpu}& 
=T_{gpu}^2\exp\Big(\frac{-\kappa_{2,gpu}}{T_{gpu}}\Big).
\end{align}
Similarly, the Jacobian $\J_\f(\Tv)$ can be written as,
\begin{multline}\label{eq:jacobian_simplified}
\J_\f(\Tv)= (\A-\I)+ p_{big,2}\tilde{s}_{big}b_{1}\begin{bmatrix}
0 & 0 & 1 & 0 & 0
\end{bmatrix} \\
+ p_{gpu,2}\tilde{s}_{gpu}b_{4}\begin{bmatrix}
0 & 0 & 0 & 0 & 1
\end{bmatrix}
\end{multline}
where $b_{i}$ denotes the $i$th column of the $\B$ matrix and
\begin{align} \nonumber
\tilde{s}_{cpu} =& \exp\Big(\frac{-\kappa_{2,big}}{T_{big}}\Big)(2T_{big} + 
\kappa_{2,big}),
\\ \tilde{s}_{gpu} =& 
\exp\Big(\frac{-\kappa_{2,gpu}}{T_{gpu}}\Big)(2T_{gpu} + 
\kappa_{2,gpu}).
\end{align}
Since $\tilde{s}_{cpu}$ and $\tilde{s}_{cpu}$ are 
scalars,~\eqref{eq:jacobian_simplified} can be written as,
\begin{align*}
\begin{split}\J_\f(\Tv)={}& (\A-\I)+ 
p_{big,2}b_{1}\tilde{s}_{big}\begin{bmatrix}
0 & 0 & 1 & 0 & 0
\end{bmatrix}
 \\ & + p_{gpu,2}b_{4}\tilde{s}_{gpu}\begin{bmatrix}
0 & 0 & 0 & 0 & 1
\end{bmatrix}, \end{split}\\
\begin{split} ={} &(\A-\I) + \begin{bmatrix}
 p_{big,2}b_{1} & p_{gpu,2}b_{4}
 \end{bmatrix}_{5\times 2}
 \begin{bmatrix}
 \tilde{s}_{big} & 0 \\
 0 & \tilde{s}_{gpu}
 \end{bmatrix}_{2\times 2} \\
 &
  \begin{bmatrix}
 0 & 0 & 1 & 0 & 0 \\
 0 & 0 & 0 & 0 & 1
 \end{bmatrix}_{2\times 5}, \end{split}\\
\begin{split} ={}& (\A-\I) \\ & + \begin{bmatrix}
 p_{big,2}b_{1} & p_{gpu,2}b_{4}
 \end{bmatrix}_{5\times 2}
 \begin{bmatrix}
 \tilde{s}_{big} & 0 \\
 0 & \tilde{s}_{gpu}
 \end{bmatrix}_{2\times 2}
 \mathbf{V}_{2\times 5}, \end{split}
\end{align*}
where $\mathbf{V} = \begin{bmatrix}
0 & 0 & 1 & 0 & 0 \\
0 & 0 & 0 & 0 & 1
\end{bmatrix}_{2\times 5}$.

\noindent When Newton's method is used to solve the fixed-point function, the 
temperature step in each iteration is given by
\begin{equation}
\Delta \Tv = -\J^{-1}_\f(\Tv)\f(\Tv).
\end{equation}
Without changing the value of $\Delta \Tv$, we can write
\begin{equation}
\Delta \Tv = -((\A-\I)^{-1}\J_\f(\Tv))^{-1}(\A-\I)^{-1}\f(\Tv).
\end{equation}
%
%
Now, let $\f_1(\Tv) = (\A-\I)^{-1}\f$ and $\tilde{\f}_1(\Tv) = 
(\A-\I)^{-1}\J_\f(\Tv)$. 
Using~\eqref{eq:f_simplified}, $\f_1(\Tv)$ is written as,
\begin{equation}\label{eq:f1}
\f_1(\Tv) = \Tv + (\A-\I)^{-1} \B \begin{bmatrix}
p_{little} \\
p_{big,1} + p_{big,2} s_{big} \\
p_{mem} \\
p_{gpu,1} + p_{gpu,2} s_{gpu}
\end{bmatrix}.
\end{equation}
Similarly, $\tilde{\f}_1(\Tv) = (\A-\I)^{-1}\J_\f(\Tv)$ is written as,
\begin{multline}\label{eq:f1_tilde}
\tilde{\f}_1(\Tv) = \I + (\A-\I)^{-1}\begin{bmatrix}
p_{big,2}b_{1} & p_{gpu,2}b_{4}
\end{bmatrix}_{5\times 2} \\
\begin{bmatrix}
\tilde{s}_{big} & 0 \\
0 & \tilde{s}_{gpu}
\end{bmatrix}_{2\times 2}
V_{2\times 5}.
\end{multline}
In order to simplify the notation, we can define matrices $\mathbf{U}$ and 
$\mathbf{C}$ as,
\begin{align}\nonumber
\mathbf{U} ={}& (\A-\I)^{-1}
\begin{bmatrix}
p_{big,2}b_{1} & p_{gpu,2}b_{4}
\end{bmatrix}, 
\\ 
\mathbf{C} ={}& (\A-\I)^{-1} \B \begin{bmatrix}
p_{little} & 0 & 0 \\
p_{big,1} & p_{big,2} & 0 \\
p_{mem} & 0 & 0 \\
p_{gpu,1} & 0 & p_{gpu,2}
\end{bmatrix}.
\end{align}
Using this notation,~\eqref{eq:f1} can be rewritten as,
\begin{equation}
{\f}_1(\Tv) = \Tv + c_1 + \begin{bmatrix}
c_{2} & c_{3}
\end{bmatrix} \begin{bmatrix}
s_{big} \\
s_{gpu}
\end{bmatrix}.
\end{equation}
where $c_i$ denotes the $i$th column of $\mathbf{C}$.
Similarly,~\eqref{eq:f1_tilde} can be rewritten as,
\begin{equation}\label{eq:f1_tilde_simplified}
\tilde{\f}_1(\Tv) = \I + \mathbf{U}\begin{bmatrix}
\tilde{s}_{big} & 0 \\
0 & \tilde{s}_{gpu}
\end{bmatrix}\mathbf{V}.
\end{equation}
In order to calculate the step size $\Delta \Tv$, the inverse of $\tilde{f}_1$ 
has to be evaluated at each step. The inverse can be easily calculated using 
the matrix inversion lemma~\cite{meyer2000matrix} on~\eqref{eq:f1_tilde_simplified}. 
Consequently, the inverse of $\tilde{\f}_1(\Tv)$ 
in~\eqref{eq:f1_tilde_simplified} is written as
\begin{equation}
\tilde{\f}^{-1}_1(\Tv) = \I - \mathbf{U}\Bigg(\begin{bmatrix}
1/\tilde{s}_{big} & 0 \\
0 & 1/\tilde{s}_{gpu}
\end{bmatrix} 
+\mathbf{VU}\Bigg)^{-1}\mathbf{V}.
\end{equation}
This simplification reduces to an inversion of a $2\times 2$ 
matrix, which can be achieved with simple algebraic operations. The evaluation 
of 
$\mathbf{U}$ requires 
computing 
$(\A-\I)^{-1}$, which can be computed offline and stored in the system.

\noindent Using~\eqref{eq:f1} and~\eqref{eq:f1_tilde_simplified}, the 
temperature step can be written as,
\begin{equation}\label{eq:delta_t_final}
\Delta{\Tv} = -\tilde{\f}^{-1}_1(\Tv)*{\f}_1(\Tv).
\end{equation}
It can be seen that the calculation of $\Delta{\Tv}$ consists of only a matrix 
multiplication and the matrix inversion is eliminated. This translates to a 
signification reduction in the computational overhead of the Newton 
iterations. A detailed analysis of the execution time savings is presented in 
Section~\ref{sec:implementation_overhead}.
\section{Temperature Control using Fixed Points} \label{sec:control_algorithm}
One of the primary applications of the fixed-point prediction is to enable 
better dynamic thermal and power management~(DTPM) algorithms. State-of-the-art 
DTPM algorithms typically throttle the entire system when a thermal violation 
is detected. This causes a performance loss for all the applications running in 
the system. Moreover, DTPM algorithms start throttling only after the 
temperature violates a threshold, which is not desirable.
In contrast to these approaches, the fixed-point prediction can be used to make
better DTPM decisions. It provides an estimate 
of the long-term thermal behavior of the system. This estimate can be used to 
determine the possibility of a thermal violation in the future. In addition to 
the fixed-point prediction, the time to reach the fixed-point 
estimate~\cite{bhat2017power} gives a lower bound on the time at which the 
fixed point is attained.

\begin{figure}[h!]
	\centering
	\includegraphics[width=0.7\linewidth]{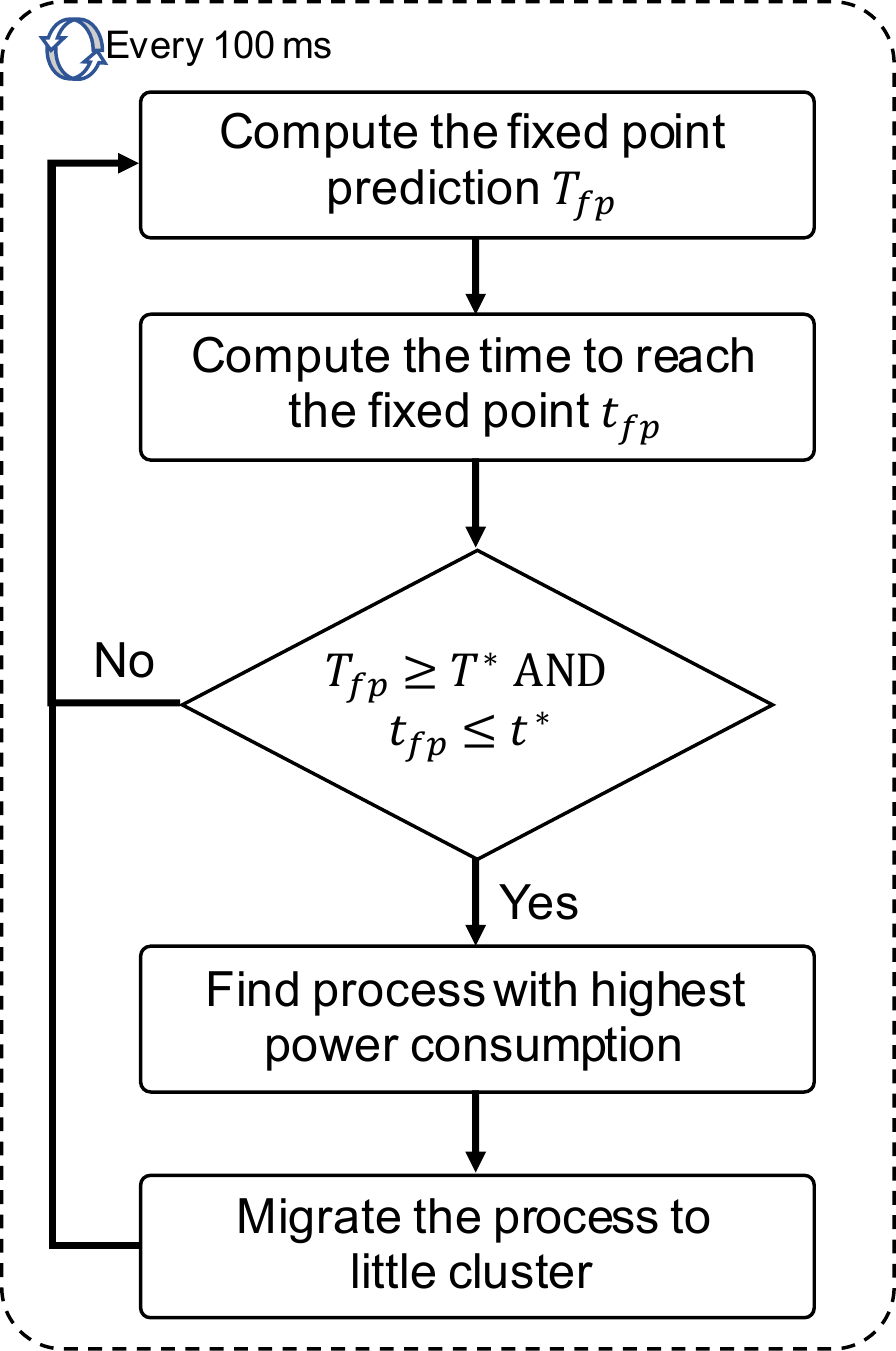}
	\caption{Overview of the proposed control algorithm.}
	\label{fig:control_algorithm}
\end{figure}

Estimates of the fixed point and the time to reach can be used to design a 
simple control algorithm, as shown in Figure~\ref{fig:control_algorithm}. The 
algorithm starts by computing the fixed point $T_{fp}$, followed by the time to 
reach the fixed point $t_{fp}$. A thermal violation is imminent if $T_{fp}$ is 
greater than a specified 
thermal limit $T^*$ and $t_{fp}$ is less than a specified time limit $t^*$.
In these cases, the algorithm finds the process with the highest power 
consumption.
This is achieved by monitoring the utilization of active processes over a 
one-second window and choosing the process with the highest 
utilization.\footnote{We can also use tools such as Powertop 
(\url{https://01.org/powertop}) to obtain precise estimates of the power 
consumption of each process.} 
We use a one-second window to ensure that momentary peaks in the power consumption are filtered out.
Then, it moves the process to the low-power processors in the 
system. Once the process is moved to the low-power processor it is kept there for a fixed period (one control interval in our implementation). After the period is over, the default scheduling algorithms on the device are free to move the process back to the big core as a function of its processing requirements. This ensures that the process is not starved of computing resources once the temperature of the device is within safe limits.
In addition to moving the process to low-power processors, we can also reduce the frequency of the core it is running on such that processes running on other cores are not penalized. However, the Odroid XU3 board used in our experimental evaluations allows control of frequency at the granularity of clusters (A15 or A7). Due to this limitation of the hardware platform, the control algorithm migrates the process with the highest power consumption to the low-power processors. 

We invoke the control algorithm every 100 ms at runtime. We choose a 100 ms interval since the frequency management governors in typical smartphone operating systems, e. g. Android, apply frequency control decisions with this period. That is, the frequency governors present in the system evaluate the utilization of resources and the workload every 100 ms. Based on this evaluation, new frequencies for each resource are set. Therefore, the 100 ms interval for the invocation of the control algorithm allows us to capture these dynamic changes in the system (e.g. a new task starts to consume high power and raises the temperature).
We note that the execution interval of the control algorithm can be changed dynamically as a function of the rate at which new processes are launched. For example, if new applications are being launched more frequently than every 100 ms we can decrease the execution interval of the control algorithm at the expense of increased runtime overhead.

The main benefit of our algorithm is that it only penalizes the processes that lead to thermal violations.
All other processes running in the 
system are not penalized; hence, they can continue to operate with maximum 
performance. Furthermore, processes with real-time requirements can register themselves,
so that they are not penalized by 
the algorithm. An illustration of the proposed algorithm is presented in the 
experiments.

One of the most important components of the control algorithm is the accurate 
prediction of the time to reach the fixed point. The approach proposed 
in~\cite{bhat2017power} uses two temperature readings separated by a fixed 
amount of time to predict the time to reach the fixed point. While this 
approach provides a prediction with a very low computational overhead, the 
error in the prediction can be high. Therefore, in this paper, we use an 
improved approach that utilizes a larger number of samples to determine 
$t_{fp}$. 
Specifically, we maintain a vector that contains the envelope of the 
temperature measurements. The envelope of the temperature at any time step $k$ 
can be expressed as:
\begin{equation}\label{eqn:envelope}
T_u[kT_s] = \max \{T[(k-M)T_s], T[(k-M+1)T_s],\dots, T[kT_s]\}
\end{equation}
where $T_u[k]$ is the envelope temperature at time $kT_s$ using a sampling 
period of $T_s$, $M$ is the size of the 
window over which we take the maximum, $T[(k-M)T_s]$ is the temperature at time 
$(k-M)T_s$ and $T[kT_s]$ is the temperature at time $kT_s$. We take the maximum 
over 
a 
window to ensure that momentary spikes in the temperature are filtered out. An 
example of the temperature envelope is provided in 
Figure~\ref{fig:fit_comparison} in the experimental evaluations.
The envelope data is then used to fit a first-order exponential model given by:
\begin{equation}\label{eqn:first_order}
T[kT_s] = T_{u,init} + (T_{fix} - T_{u,init})(1 - e^{-\frac{kT_s}{\tau}})
\end{equation}
where $T_{u,init}$ is the initial upper envelope temperature, $T_{fix}$ is the 
fixed-point prediction and $\tau$ is the time constant for the first-order 
model. We use a nonlinear curve fitting tool to fit the upper envelope 
temperature data to the model in~\eqref{eqn:first_order}. The quality of the 
fit depends on the number of data points used in the curve fitting. A larger 
number of data points ensures a fit with lower error. However, the system needs 
to wait for a longer period of time to allow the accumulation of upper 
envelope temperature. In contrast, fewer data points can provide a faster 
estimation of the time to fixed point, albeit with a higher error. We explore 
this trade-off to determine the optimal number of data points, as shown in 
Section~\ref{sec:time_to_fp}. In addition to the first-order exponential model, 
we evaluated 
a second-order model to estimate the time to the fixed point. However, we 
observed that the first time constant dominates the second time constant, 
implying that the model is strongly first order. Therefore, we use a 
first-order model in our experimental evaluations.
\section{Experimental Evaluation} \label{sec:experiments}
\subsection{Experimental Setup} \label{sec:experimental_setup}
We perform evaluations of the proposed fixed-point prediction scheme using the 
Odroid-XU3 board. The Odroid-XU3 board uses the Samsung Exynos 5422 
system-on-chip~\cite{ODROID_Platforms} that integrates four Cortex-A15~(big) 
cores, four Cortex-A7~(little) cores, and a Mali T628 
GPU. It also includes thermal sensors 
to measure the temperature of each big core and the GPU. Furthermore, the 
platform provides 
current sensors that allow us to obtain the power consumption of the little cluster, big 
cluster, main memory, and the GPU at runtime. The board is installed with 
Android 7.1 
running Linux kernel 3.10.9. We invoke the fixed-point framework and the control algorithm every 100 ms, along with the default frequency governors. 
The proposed approach is evaluated on Android benchmarks, such as 3DMark and Nenamark3. 
We choose 3DMark and Nenamark3 benchmarks because they are able to give standard metrics to 
measure the performance of various devices and algorithms. 
Both benchmarks run a series of tests on the device and report a metric at the end. 
Hence, we can compare the performance of the Odroid-XU3 board with and without our 
proposed approach.

\subsection{Parameter Estimation and Accuracy Evaluation} \label{sec:sys_id}
The thermal fixed-point analysis presented in this paper does not assume any 
specific values for the system parameters. However, in order to validate the 
fixed-point analysis, system parameters have to be characterized. 
Specifically, we characterize the system parameters (e.g., matrices $\A$, $\B$, 
and the leakage power)
for the experimental platform. 
We use the \textit{ssest} function in the system identification of MATLAB~\cite{matlab} to the thermal model matrices $\A$ and $\B$. From the system identification, we observe that both $\A$ and $\B$ matrices are full rank.
Similarly, nonlinear curve fitting is used to estimate the leakage power parameters for the big cluster and GPU. 
Further details on the methodology used to characterize the parameters are presented in~\cite{bhat2017power} and~\cite{bhat2017algorithmic}.

\vspace{1mm}
\noindent\textbf{Accuracy of Fixed-Point Predictions:} The effectiveness of the proposed control algorithm depends on the accuracy of the fixed-point predictions. Therefore, we perform a detailed accuracy evaluation with ten compute-intensive benchmarks.
Across the ten benchmarks, the highest error observed is 5.8$\degree$C which translates to a 6\% error. On average, the proposed analysis predicts the fixed point with about 3$\degree$C~(5\%) error. For more details of the accuracy, such as the error for each benchmark, please refer to~\cite{bhat2017power}.

\subsection{Power-Temperature Trajectory for MIMO Iterations}

\begin{figure}[h]
	\centering
	\includegraphics[width=0.95\linewidth]{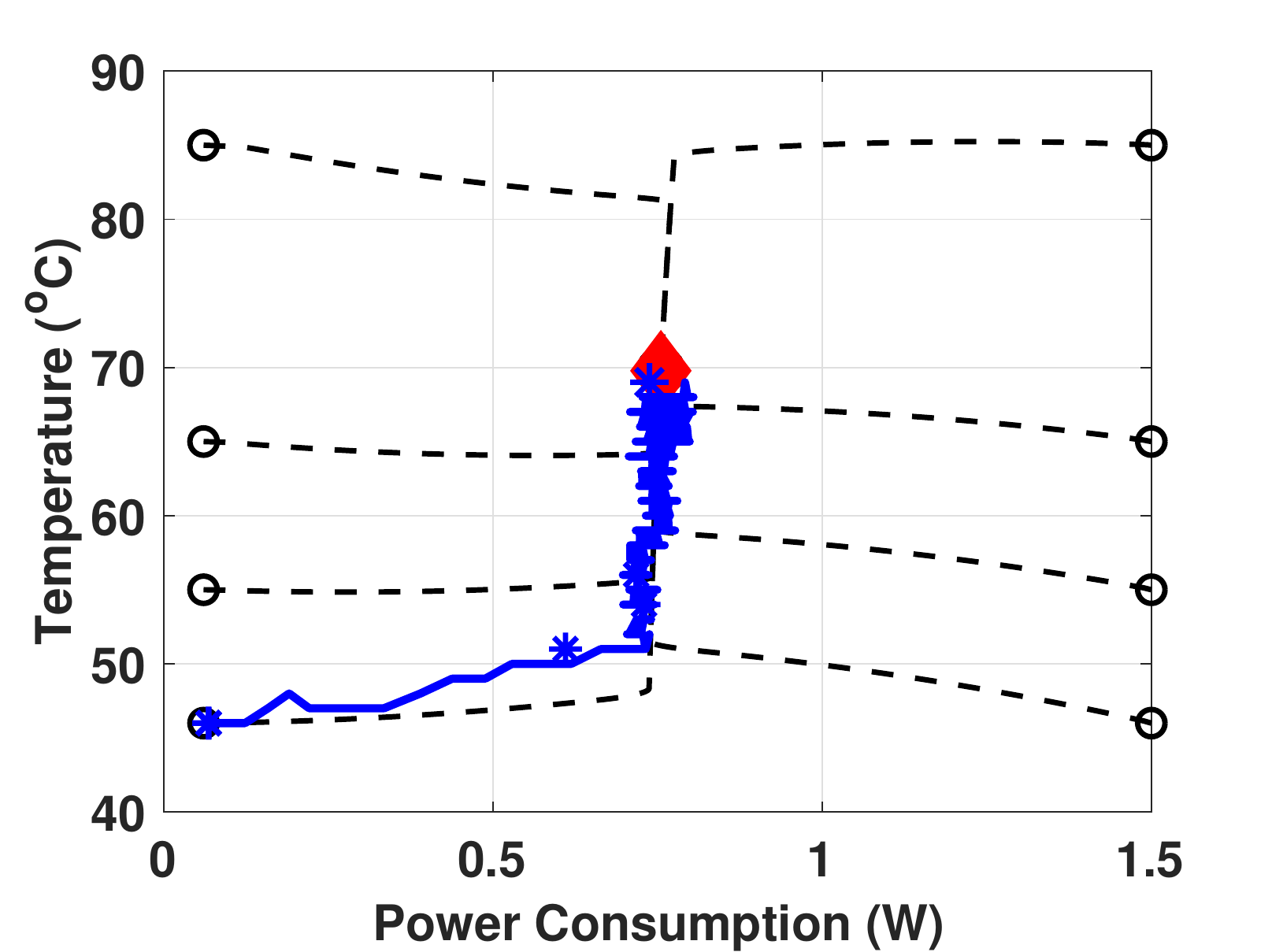}
	\vspace{-0.05in}
	\caption{Power trajectory of the GPU when the total power consumption of the device is held constant at about 1.18 W}
	\label{fig:pt_trajectory_gpu}
	\vspace{-5mm}
\end{figure}

In this section, the fixed-point prediction is compared against the measured 
power-temperature trajectory for the GPU. In order to get an accurate 
experimental fixed 
point, the total power consumption of the device is kept at a constant level of about 1.18 W. Then, the fixed point of 
the GPU is calculated using the measured power consumption values, as shown 
using a red diamond in Figure~\ref{fig:pt_trajectory_gpu}. It is seen that the 
experimental fixed-point value closely matches the predicted fixed point. In 
addition 
to the prediction, simulations using different initial values of temperature 
and power consumption are performed. The simulations use the thermal and 
leakage power models to update the temperature and leakage power iteratively. 
We see that all the simulated trajectories converge to the predicted 
fixed point. Moreover, the measured power-temperature trajectory closely 
follows the simulated trajectory. This demonstrates that the thermal and power 
models identified closely match the dynamics of the device.

The temperature trajectory in Figure~\ref{fig:pt_trajectory_gpu} can also be used to estimate CPU to environment thermal resistance of the device. The thermal resistance is defined as:
\begin{equation}
    \theta = \frac{\Delta T}{P}
\end{equation}
where $\theta$ is the thermal resistance, $\Delta T$ is the temperature differential, and $P$ is the power consumption. From the figure, we see that $\Delta T$ is 23$\degree$C while the power consumption $P$ is set as 1.18 W. Using this the thermal resistance $\theta$ is obtained as 19.48$\degree$C/W. We can also obtain the thermal resistance using the SISO model in~\eqref{eqn:temp_ss}. Specifically, we can write the thermal resistance as:

\begin{equation}
    \theta = \frac{b}{1-a}.
\end{equation}
Using the parameters obtained using system identification for $a = 0.9994$ and $b = 0.0121$, we can calculate the thermal resistance as 20.1$\degree$C/W, which is in close agreement to the value obtained experimentally. This shows that the proposed system identification methodology is able to accurately capture the thermal dynamics of the system. We note that more detailed models for the thermal resistance can be obtained if the details about the location of temperature sensors and chip packaging are publicly available. However, in the absence of these details we use system identification to gain insights into the thermal dynamics of the system.

\subsection{Time to Reach the Fixed Point}\label{sec:time_to_fp}
\begin{figure}[t]
	\centering
	\includegraphics[width=1\linewidth]{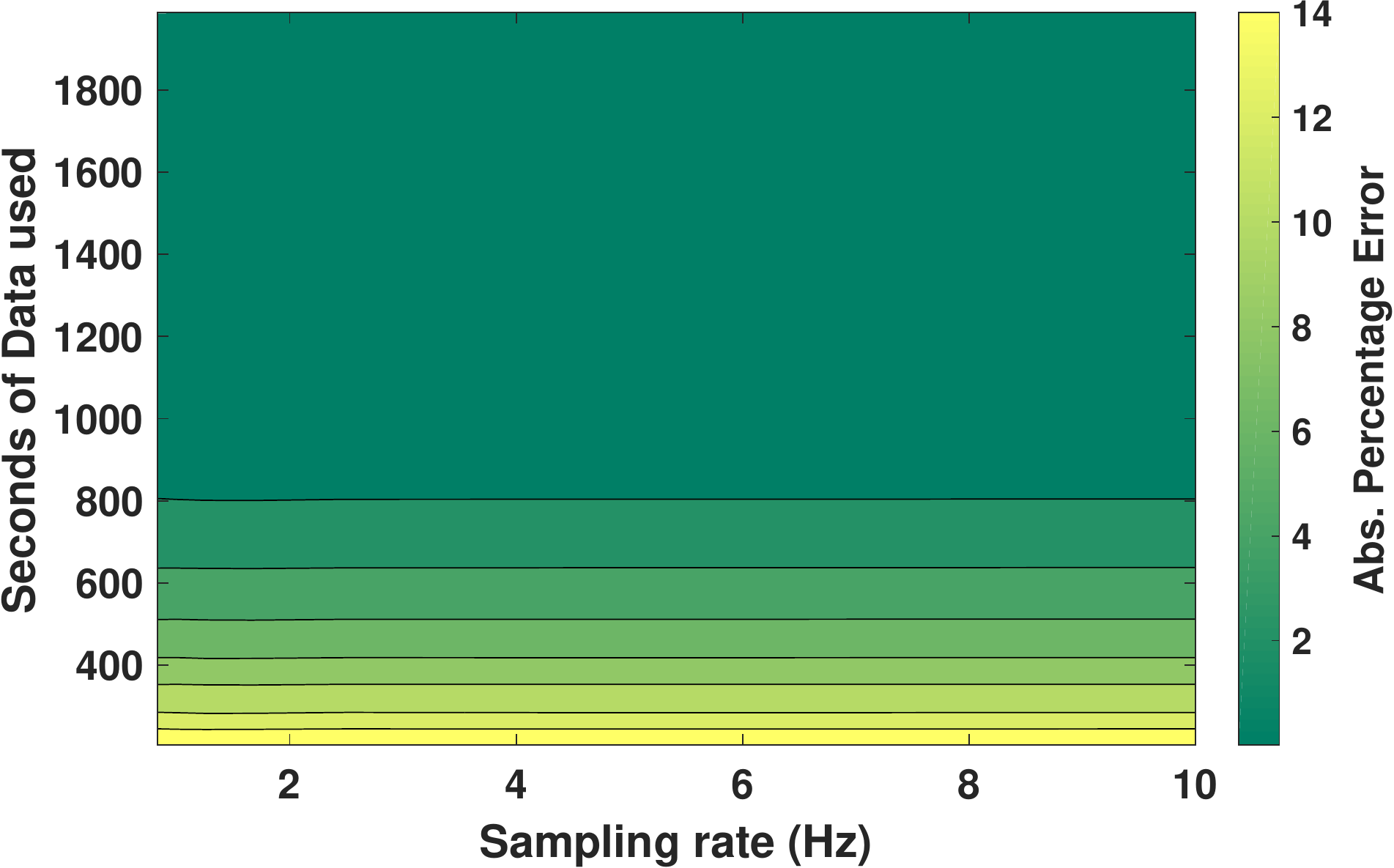}
	\caption{Comparison of prediction of time to the fixed point.}
	\label{fig:tau_comparison}
\end{figure}

Accurate estimation of the time to reach the fixed point is an important 
component of the proposed control algorithm. Underestimation of the time to 
fixed point can lead to thermal issues, while overestimation leads to 
performance loss due to aggressive throttling.
As explained in Section~\ref{sec:control_algorithm}, 
the accuracy of the time to fixed-point estimation depends on the number of data
points used to fit the first-order model. Therefore, we evaluate the 
accuracy of the time to fixed-point estimation as a function of the number of 
data points used for the estimation. We can vary the number of samples used 
in the fitting by controlling the delay in fitting or by controlling the 
sampling rate of the temperature measurements.
Figure~\ref{fig:tau_comparison} shows the error in the time to the fixed point as we vary 
the sampling rate and the delay in fitting. 
We observe that using temperature measurements in a shorter window and 
a lower sampling rate results in a higher error. 
As we increase the length of the window and the sampling rate, the error progressively reduces. 
We also observe that length of the window is more important than the 
sampling rate of the temperature measurements. 
For instance, using a 200~s window at a sampling rate of 10 Hz results in an 
error of only 5\%, while shorter windows result in a higher error. A 200~s 
window is acceptable to obtain an accurate 
estimation of the time to fixed point as it takes more than 500~s for the system to attain steady state.

\begin{figure}[h]
	\centering	
	\includegraphics[width=1\linewidth]{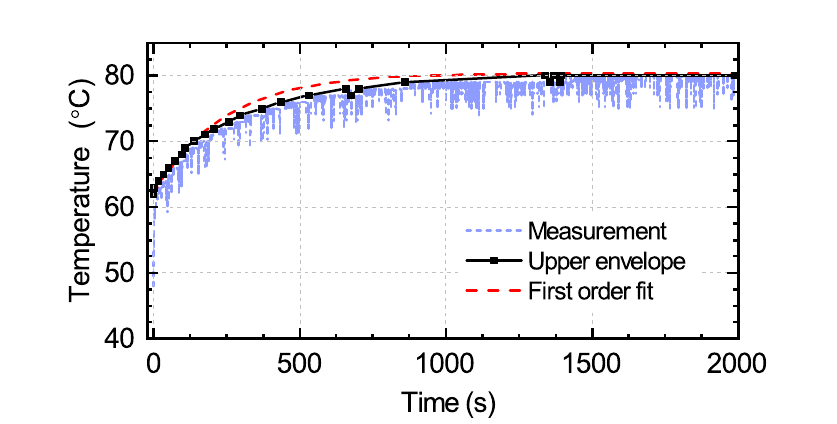}
	\vspace{-5mm}
	\caption{Comparison of temperature measurement, upper envelope of the temperature, and our first-order fit.}
	\label{fig:fit_comparison}
\end{figure}

Figure~\ref{fig:fit_comparison} shows the comparison of the measured 
temperature, upper envelope 
temperature, and the first-order fit, for the Vortex workload in the SPEC 
benchmark suite, using the temperature measurements in the 
first 200~s. We see that the first-order fit closely follows the upper envelope 
temperature. Moreover, the time at which the first-order fit reaches the steady 
state differs from the actual time by about \textcolor{black}{10\%}.
This error is acceptable since the fit is repeated every second. Therefore, as 
more data becomes available, the error decreases continuously. As a result of 
this, the control algorithm has the most up-to-date estimation of the 
time to the fixed point.

\subsection{Application Control Using Fixed-Point predictions}
This section presents an illustration of the control algorithm using 
fixed-point predictions. We run a real-time GPU 
benchmark along with a computationally intensive task in the background to 
evaluate the effectiveness of the control algorithm.
The default policy is to use the thermal management algorithm in the Linux kernel version 3.10.9. 
Specifically, it uses fixed temperature thresholds and the ARM intelligent 
power allocation~(IPA) algorithm to control the
temperature~\cite{armIPA}. ARM IPA uses a proportional-integral-derivative feedback controller (PID) to manage the temperature of the device. The IPA policy has a control temperature set at the time of its initialization. At runtime, it compares the control temperature to the maximum temperature observed in the system. Then, it finds the error between the control temperature and measured maximum temperature. Based on this error term, the IPA algorithm sets the frequency of each computing resource. We use the IPA algorithm as the baseline in our comparisons.
The on-board fan is disabled during these evaluations 
since it is not a viable option for mobile platforms.

\begin{table}[t]

\caption{Maximum temperature of the system when using the default policy, 
		DTPM, and the proposed control algorithm, respectively. A temperature violation occurs when the maximum temperature is higher than 85$\degree$C}
		\label{tab:violations}
\centering
\begin{tabular}{@{}lc@{}}
\toprule
Scenario & Number of thermal violations \\ \midrule
3DMark (Baseline) & 19 \\
3DMark + BML (Baseline) & 1294 \\
3DMark + BML (Proposed approach) & 176 \\ \bottomrule
\end{tabular}
\end{table}

\begin{table}[b]
	\centering
	\vspace{-5mm}
	\caption{Comparison of performance of three applications with proposed 
	control}
	\label{table:perf_comparison}
	\begin{tabular}{@{}lccc@{}}
		\toprule
		Test                & App. Alone & App. + BML & 
		\begin{tabular}[c]{@{}c@{}}App. + BML with\\ Proposed 
			Control\end{tabular} \\ \midrule
		3DMark GT1          & 97 fps   & 86 fps   & 93 
		fps                                                                   \\
		\midrule
		3DMark GT2          & 51 fps   & 49 fps   & 52 
		fps                                                                   \\
		\midrule
		Nenamark3           & 3.5 levels & 3.4 levels & 3.5 
		levels                                                                 
		\\
		\bottomrule
	\end{tabular}
\end{table}

\begin{figure}[t]
	\centering
	\includegraphics[width=0.95\linewidth]{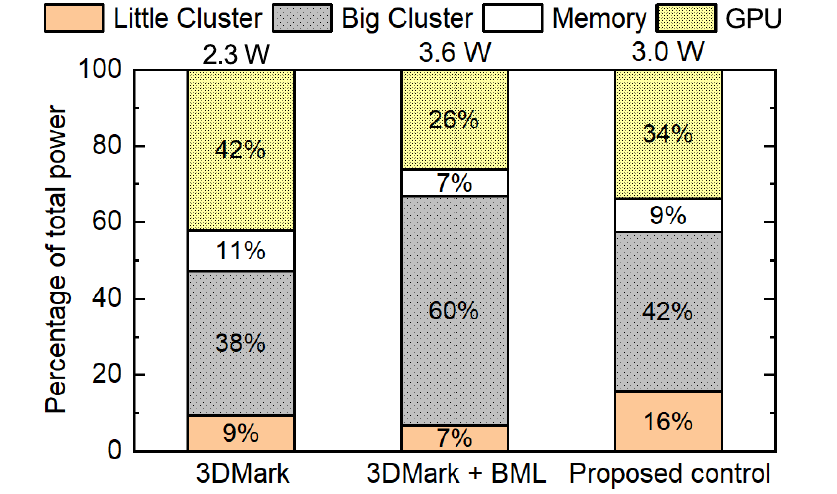}
	\vspace{-3mm}
	\caption{Power consumption distribution of 3DMark. }
	\label{fig:3dmark_power_distrubition}
	\vspace{-5mm}
\end{figure}

\noindent \textbf{3D Mark alone:} 
We start by running the 3DMark benchmark alone without introducing any background applications.
Specifically, we run graphics test 1~(GT1) and graphics test 2~(GT2) benchmarks 
in the 3DMark application. 
We obtain an upper bound on the performance with this experiment.
This experiment also gives us the baseline temperature profile, as shown in 
Table~\ref{tab:violations}. A thermal violation occurs whenever the maximum 
temperature is greater than 85$\degree$C, the temperature at which the default 
governor starts throttling the system. The first row of 
Table~\ref{tab:violations} shows that there are 19 thermal violations when 
running 3DMark alone.
We also analyze the power consumption of each SoC component when running the 
3DMark application in
Figure~\ref{fig:3dmark_power_distrubition}.
The GPU consumes about 42\% of the total power since 3DMark is a GPU-intensive 
application. This is followed by the big cluster that consumes 38\% of the 
power to perform computational parts of 3DMark.

\vspace{1mm}
\noindent \textbf{3D Mark + BML (IPA):} 
In the second part of the evaluation, we re-run the 3DMark benchmark along with 
the Basicmath 
Large~(BML) application~\cite{guthaus2001mibench} in the background. 
BML is a CPU-intensive application that performs mathematical computations. As 
a result, it leads to an increase in the big core power consumption, which 
causes the total power consumption to increase  to 3.65~W. The increase in the 
big core power consumption also increases its contribution to the total power 
from 38\% to 60\%, as shown in Figure~\ref{fig:3dmark_power_distrubition}. The 
higher power consumption causes the temperature to increase quickly beyond 
85$\degree$C. Indeed, the number of thermal violations increases to 1294. As a 
result of this increase, the default IPA thermal governor starts to reduce the 
frequency of all the resources in the system. This leads to an undesirable drop 
in performance of GT1 and GT2,
as shown in the third column of Table~\ref{table:perf_comparison}. Specifically, the frame rate of GT1 decreases from 97 to 86 frames per second~(fps), while the frame rate of GT2 decreases 
from 51~fps to 49~fps.


\vspace{2mm}
\noindent \textbf{Proposed Control:} 
Finally, we apply the proposed control algorithm when both 3D Mark and 
BML are running on the board. As expected, BML causes the power consumption and the temperature to rise. The proposed algorithm keeps track of the increase in the power consumption and updates its estimates of the fixed point continuously. As soon as it detects that the thermal limit will be breached in the near future, it migrates BML to the little cluster to reduce the number of thermal violations. The migration successfully reduces the number violations to 176, as shown in the third row of Table~\ref{tab:violations}. The migration also leads to a reduction of the big core power contribution from 60\% to 42\%, as shown in Figure~\ref{fig:3dmark_power_distrubition}.
Moreover, the migration of BML to the little cluster causes its power 
consumption to increase from 7\% to 16\% of the total power. The migration 
effectively throttles BML with minimal performance loss for the 3DMark 
application, as seen in last column of Table~\ref{table:perf_comparison}.

\begin{figure}[t]
	\includegraphics[width=0.95\linewidth]{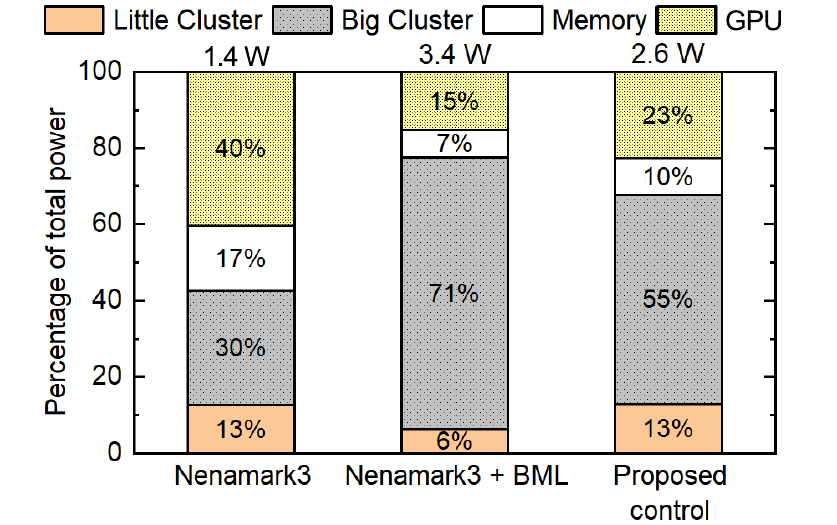}
	\caption{Power consumption distribution for the Nenamark3 benchmark.}
	\label{fig:nenamark_power_distrubiton}
	\vspace{-3mm}
\end{figure}

We repeat the same with the Nenamark3 benchmark in order to evaluate the 
control algorithm
on a benchmark with different characteristics.
The Nenamark3 benchmark measures the number of levels the platform can run at a given frame rate. 
Once the frame rate drops below the reference level, the benchmark terminates and reports the number of levels completed.
The last row of Table~\ref{table:perf_comparison} summarizes the performance obtained for the Nenamark3 benchmark under the three scenarios. As expected, the performance drops when we run BML along with Nenamark3. 
In contrast, the proposed control algorithm recovers the performance back to the baseline level. Furthermore, the power consumption follows a trend 
similar to that of 3DMark, as shown in Figure~\ref{fig:nenamark_power_distrubiton}. 
In summary, the control algorithm effectively detects and migrates the 
power-hungry applications without affecting the performance of the foreground 
application.


\begin{figure}[t]
	\centering
	\includegraphics[width=1\linewidth]{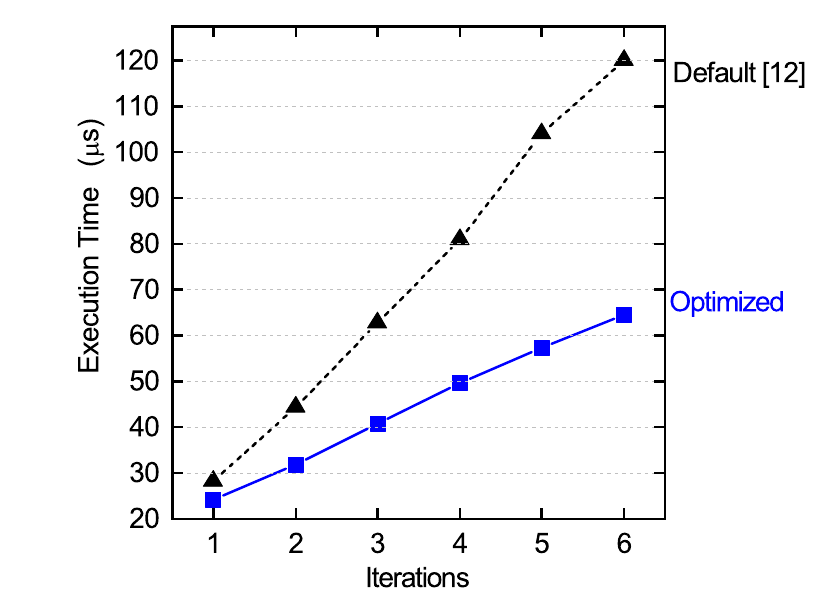}
	\caption{Comparison of the execution time of Newton's method using the 
		default implementation~\cite{bhat2017power} and the proposed optimized 
		implementation.}
	\label{fig:newton_timing}
	\vspace{-0.07in}
\end{figure}
\vspace{1mm}
\noindent\textbf{Implementation Overhead of Proposed Control:} The implementation overhead of the proposed algorithm consists of two parts. The first part involves the evaluation of the fixed points by performing Newton's method iterations. With our accelerated approach, the computation of the fixed points with six Newton iterations takes about 63 $\mu$s, as detailed in the next section. This computation has to be performed every 100~ms i.e. with every invocation of the control algorithm. The second part of the overhead is incurred when we find the process with the highest power consumption. This part takes about 1~s since we profile the utilization of all processes in the system in a 1~s window. A 1~s window is needed to ensure that momentary peaks in the 
power consumption can be filtered out. This overhead is not significant since 
it is invoked infrequently, i.e. \textit{only} when a thermal violation is 
predicted. Furthermore, we can reduce the 1~s overhead by running a background 
task that keeps track of the process with the highest power consumption. In 
summary, the control algorithm takes only about 63 $\mu$s during normal course 
of execution, which is less than 0.1\% of the 100 ms interval.

\subsection{Effect of Accelerated Fixed-Point 
Computation}\label{sec:implementation_overhead}
The runtime nature of the fixed-point computation necessitates fast and 
efficient 
implementation of the fixed-point iterations. 
Therefore, this section compares the implementation overheads of Newton's 
method with and without the proposed optimization. The default implementation 
of Newton's method takes about 27 $\mu$s for a single iteration and 
increases linearly with increasing iterations, as shown in 
Figure~\ref{fig:newton_timing}. The optimized method, on the other hand, takes 
about 24~$\mu$s for one iteration and increases at a much lower rate than the 
default implementation. Overall, we achieve speed up of about 1.8$\times$ when compared to the default implementation.

\section{Conclusion} \label{sec:conclusion}
This paper presented a theoretical analysis of the temperature dynamics in 
multiprocessor systems. We first solved the system of MIMO equations to find 
the 
temperature fixed points in the system. Then, we experimentally derived the 
region of convergence of the CPU-GPU dynamics. Finally, we 
utilized a control algorithm to manage the temperature of the system without 
sacrificing the performance. We demonstrated the control algorithm on the 
Exynos 
5422 system on a chip.

Even though we focused our analysis on mobile systems, it can be applied to 
other low-power devices such as wearables.
This can be achieved by modeling the dynamics of these systems and 
then using the MIMO solution presented in this work. The analysis can be used to improve the thermal behavior in future 
processors. In particular, the 
fixed-point analysis can be used to detect power attacks in the system by 
detecting processes that lead to a high steady-state temperature. 

\vspace{2mm}
\noindent \textbf{Acknowledgements:} 
This work was supported partially by Semiconductor Research Corporation (SRC) task 2721.001 and National Science Foundation grant CNS-1526562.


{\footnotesize

}
\begin{IEEEbiography}[{\includegraphics[width=1in,height=1.25in,clip,keepaspectratio]{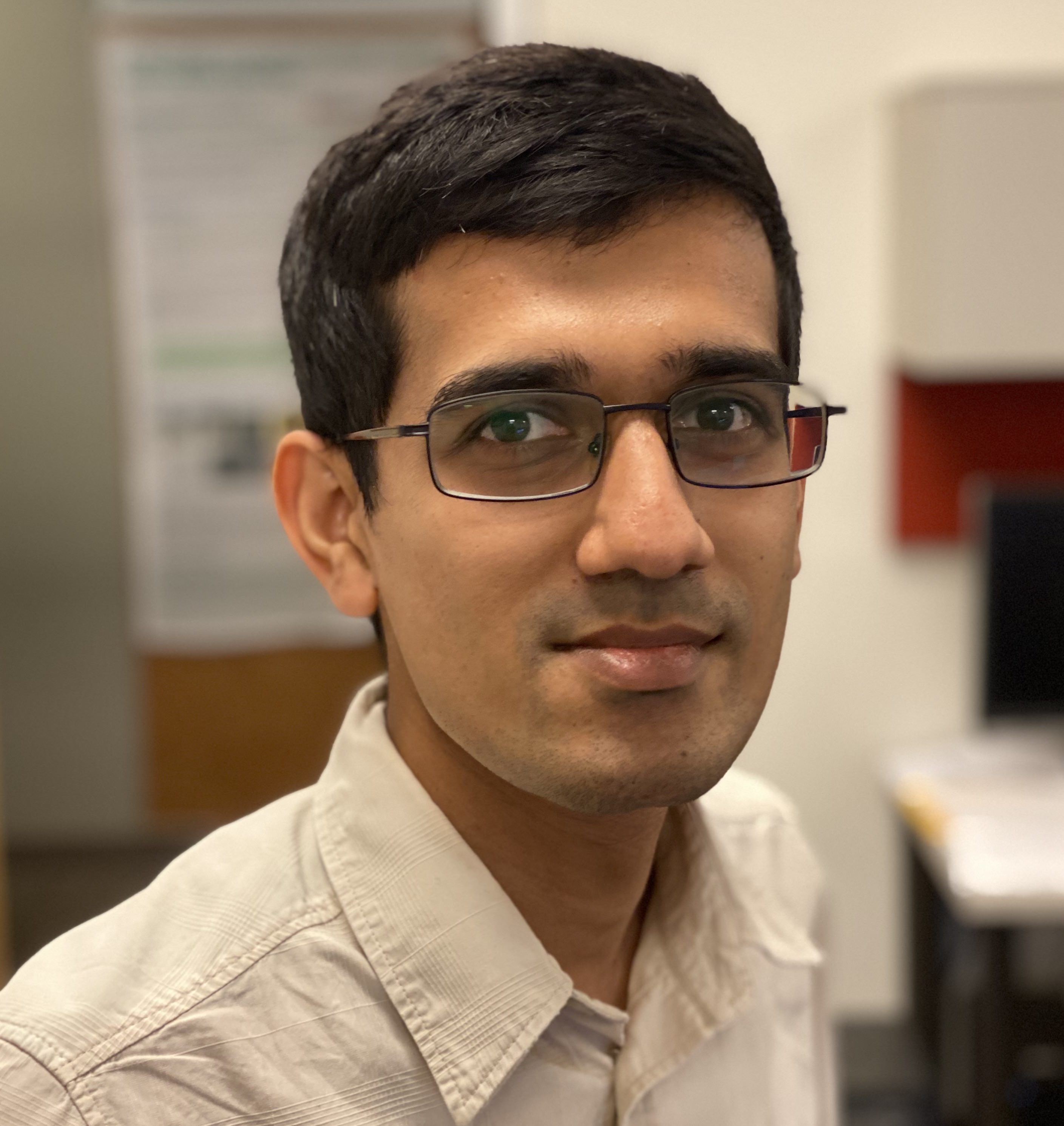}}]
		{Ganapati Bhat} received his B.Tech degree in Electronics and 
		Communication 
	from Indian Institute of Technology~(ISM), Dhanbad, India in 2012. From 
	2012-2014, he 
	worked as a software engineer at Samsung Research and Development 
	Institute, Bangalore, India. He is currently a PhD candidate in Computer 
	Engineering at the School of Electrical, Computer and Energy 
	Engineering, Arizona State University. His research interests include 
	energy optimization in computing systems, dynamic thermal and power 
	management, and energy management for wearable systems.
\end{IEEEbiography}
\begin{IEEEbiography}[{\includegraphics[height=1.25in,clip,keepaspectratio]{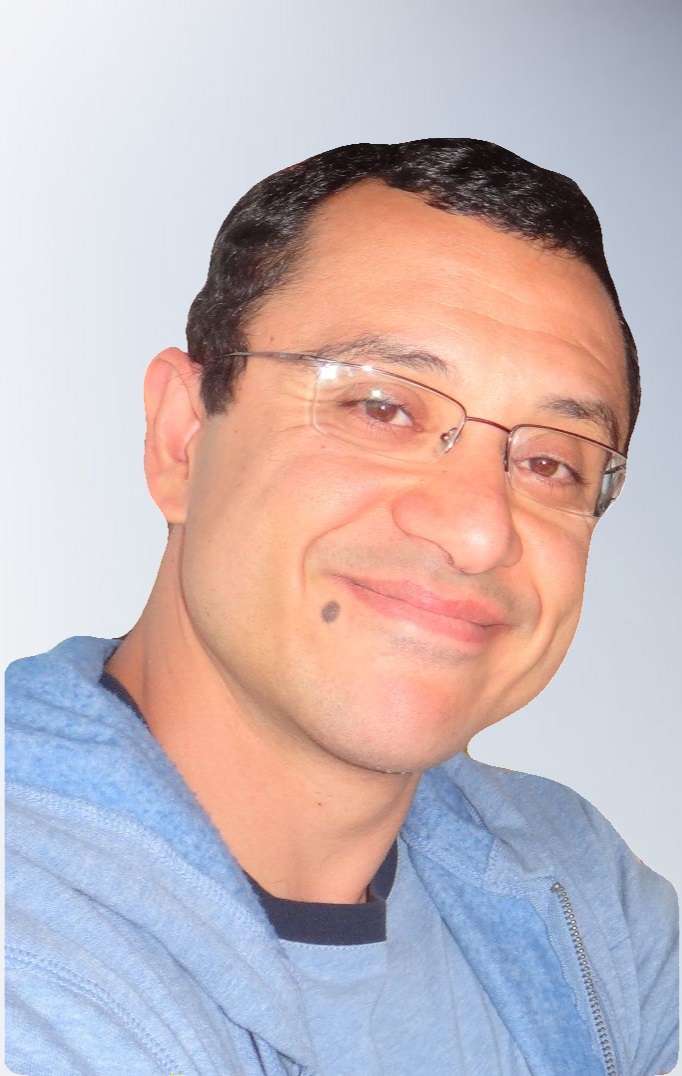}}]
{Dr. Gumussoy} is a research scientist at Autonomous Systems \& Control group at Siemens Corporate Technology in Princeton, NJ. His general research interests are learning, control, identification, optimization and scientific computing with particular focus on reinforcement learning, optimal-adaptive control, frequency domain system identification, time-delay systems and their numerical implementations. 

He serves as an Associate Editor in IEEE Transactions on Control Systems Technology and IEEE Conference Editorial Board.

Dr. Gumussoy received his B.S. degrees in Electrical \& Electronics Engineering and Mathematics from Middle East Technical University at Turkey in 1999 and M.S., Ph.D. degrees in Electrical and Computer Engineering from The Ohio State University at USA in 2001 and 2004. He worked as a system engineer in defense industry (2005-2007) and he was a postdoctoral associate in Computer Science Department at University of Leuven (2008-2011). He was a principal control system engineer in Controls \& Identification Team at MathWorks where his contributions ranges from state-of-the-art numerical algorithms to comprehensive analysis \& design tools in Control System, Robust Control, System Identification and Reinforcement Learning Toolboxes. 	
\end{IEEEbiography}
\begin{IEEEbiography}[{\includegraphics[width=1in,height=1.25in,clip,keepaspectratio]{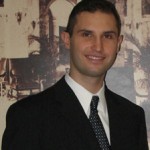}}]
{Umit Y. Ogras} received his Ph.D. degree in Electrical 
and Computer 
Engineering from Carnegie Mellon University, Pittsburgh, PA, in 2007. From 
2008 to 2013, he worked as a Research Scientist at the Strategic CAD 
Laboratories, 
Intel Corporation. He is currently an Associate Professor at the School of 
Electrical, Computer and Energy Engineering. Recognitions Dr. Ogras has
received include Strategic CAD Labs Research Award, 2012 IEEE Donald O.
Pederson Transactions on CAD Best Paper Award, 2011 IEEE VLSI Transactions 
Best 
Paper Award and 2008 EDAA Outstanding PhD. Dissertation Award. His research 
interests include digital system design, embedded systems, multicore 
architecture and electronic design automation with particular emphasis on 
multiprocessor systems-on-chip (MPSoC).
\end{IEEEbiography}

\end{document}